\documentclass[twocolumn,showpacs,preprintnumbers,amsmath,amssymb]{revtex4}
\usepackage{graphicx}% Include figure files
\usepackage{dcolumn}% Align table columns on decimal point
\usepackage{bm}% bold math

\begin{document}

\preprint{APS/123-QED}

% Titre********************************************************************************************

\title{Effect of annealing on the superconducting properties of a-Nb$_x$Si$_{1-x}$ thin films}% Force line breaks with \\

%**************************************************************************************************

% Auteurs & Adresses ******************************************************************************
\author{O. Crauste}
\author{A. Gentils}
\author{F. Cou\"{e}do}
\author{Y. Dolgorouky}
\author{L. Berg\'{e}}
\author{S. Collin}
\author{C.A. Marrache-Kikuchi}
\email{Claire.Marrache@csnsm.in2p3.fr}
\author{L. Dumoulin}
\affiliation{%
CSNSM, CNRS-UMR8609, Universit\'{e} Paris Sud, Bat. 108, 91405 Orsay Campus, France
}%

%**************************************************************************************************

% Date ********************************************************************************************
\date{\today}% It is always \today, today,

% Abstract ****************************************************************************************

\begin{abstract}
a-Nb$_x$Si$_{1-x}$ thin films with thicknesses down to 25 {\AA} have been structurally characterized by TEM (Transmission Electron Microscopy) measurements. As-deposited or annealed films are shown to be continuous and homogeneous in composition and thickness, up to an annealing temperature of 500$\,^{\circ}{\rm C}$. We have carried out low temperature transport measurements on these films close to the superconductor-to-insulator transition (SIT), and shown a qualitative difference between the effect of annealing or  composition, and a reduction of the film thickness on the superconducting properties of a-NbSi. These results question the pertinence of the sheet resistance $R_\square$ as the relevant parameter to describe the SIT.
\end{abstract}

%**************************************************************************************************

% PACS ********************************************************************************************

%**************************************************************************************************

% Keywords ****************************************************************************************

%Use showkeys class option if keyword

% Article *****************************************************************************************

\maketitle

%?????????????????????????????????????????????????????????????????????????????????????
\section{Introduction}
%?????????????????????????????????????????????????????????????????????????????????????

The interplay between superconductivity and disorder is a long-standing problem \cite{Belitz1987, Imry1981, Maekawa1982, Fisher1990, Shahar1992, Sacepe2011}. One of the first steps towards its understanding was given by the so-called Anderson theorem \cite{Anderson1959}, which states that weak non-magnetic disorder has no effect on the superconducting critical temperature $T_c$. However, this limit is rarely ever attained in real samples. Moreover, numerous experiments have subsequently shown that microscopic interactions relate to the level of disorder in a more complex manner \cite{Suhl1980, Strongin1970}. Despite a few decades of both theoretical and experimental efforts, this problem remains unsolved and is still a major issue in solid state physics \cite{Feigelman2010}.

The effects of disorder on superconductivity are all the more dramatic in 2D, which is the lowest dimension for the existence of either a metal or a superconductor. In 2D, as the microscopic disorder is increased, superconducting thin films evolve towards an insulating state. This change in ground state has commonly been described as a Superconductor–-to--Insulator Transition (SIT) and results from the competition between disorder-induced localization of charge carriers and the formation of a coherent state of condensed Cooper pairs \cite{Sondhi1997,Feigelman2010}. Within these theoretical descriptions, the physical properties of the films should only depend on the sheet resistance $R_\square\propto \frac{1}{k_Fl}$ \cite{Bergmann1976, Shahar1992} where $k_F$ is the Fermi wave vector and $l$ the mean free path.

The disorder-induced SIT has been studied experimentally in several systems \cite{Deutscher1980, Jaeger1989, Markovic1998, Lee1990, Shahar1992, Bielejec2002, Hadacek2004}. In these studies, the experimental tuning parameter is usually a variation in the sample thickness. How exactly a reduction in sample thickness affects $R_\square$ is however unclear: although both processes affect $R_\square$, a change in the film microscopic disorder and a reduction of the sample thickness might be different in nature. Varying the sample thickness around typical values of a few tens of angstr\"{o}ms may change the relative influence of the surface, as has been previously suggested \cite{Keck1976, Montroll1950}, or even modify the effective dimensionality of the system. The study of the influence of microscopic disorder on superconducting properties in samples of the \emph{same} thickness might therefore help disentangle the complex effects gathered under the generic term of "disorder".

Our system of interest is amorphous Nb$_x$Si$_{1-x}$ (a-NbSi) which is known to undergo a disorder-driven SIT for 2D samples where the film thickness $d<\xi_{SC}$, the superconducting coherence length \cite{Marrache2008}. The aim of this paper is to investigate the effect of annealing (up to 250$\,^{\circ}{\rm C}$), as another way to vary $R_\square$, on the SIT. With this approach, annealing is a new parameter that allows us to microscopically finely modify the film disorder within the same sample, without changing its thickness. We will first show that these moderate thermal treatments do not measurably affect the structure of the samples (Sec. III). We will then focus on the modifications thus induced on their superconducting properties (Sec. IV). Section V will aim at providing a possible interpretation of this effect.

%?????????????????????????????????????????????????????????????????????????????????????

%?????????????????????????????????????????????????????????????????????????????????????
\section{Experimental procedure}
\label{sec:synthesis}
%?????????????????????????????????????????????????????????????????????????????????????

All a-NbSi films have been prepared at room temperature and under ultrahigh vacuum (typically a few 10$^{-8}$ mbar) by electron beam co-deposition of Nb and Si, at a rate of the order of 1 {\AA}.s$^{-1}$. We have shown special care in the control of the sample parameters: the evaporation was controlled in situ by a special set of piezo-electric quartz in order to precisely monitor the composition and the thickness of the films during the deposition. In order to check ex situ these two parameters, Rutherford Backscattering Spectroscopy (RBS) \cite{Walls} measurements were systematically undertaken and the results well fitted with the in situ monitoring. Due to heating during the deposition process, the equivalent annealing temperature of the as-deposited films has been estimated at 70$\,^{\circ}{\rm C}$, and confirmed by low temperature measurements of the conductivity (see section \ref{subsec:conductivity}).

Films for \emph{transport measurements} were deposited onto sapphire substrates coated with a 250 {\AA}-thick SiO underlayer designed to smooth the substrate surface. The samples were subsequently protected from oxydation by a 250 {\AA}-thick SiO overlayer. The Nb concentrations ranged from 13.5\% to 18\% and the film thicknesses from 20 to 500 {\AA}. Each film has been annealed, by steps, from 70$\,^{\circ}{\rm C}$ to 250$\,^{\circ}{\rm C}$, for 1 hour, under a flowing N$_2$ atmosphere. To prevent any thermal shock, the samples were slowly cooled to room temperature.

Atomic Force Microscopy (AFM) and Transmission Electron Microscopy (TEM) measurements have been carried out on Nb$_{18}$Si$_{82}$ films of thicknesses down to 25 {\AA}, close to the lowest sample thickness studied by transport measurements (20 {\AA}). Although the transport samples were of different compositions, it is reasonable to assume that the film stoichiometry does not affect its structure, at least in the considered range of niobium concentrations (13.5\%$<x<$18\%).

Samples for \emph{AFM measurements} were deposited onto a silicon wafer, overcoated with a native silicon oxide layer. The a-NbSi samples were not covered by a SiO overlayer, in order to accurately characterize the film morphology.

\emph{TEM studies} were performed on films evaporated onto commercial 250 {\AA}-thick SiO$_2$ membranes, using a Tecnai G$^2$ FEI microscope operated at 200 keV. EFTEM (Energy Filtered Transmission Electron Microscopy) measurements have also been performed on this microscope, using a GATAN Tridiem imaging filter. As-deposited a-NbSi films were only covered by their native oxyde layer. In order to assess the influence of the annealing temperature on the film structure, samples capped with 100 {\AA} of SiO have been synthesized and embedded within a heating TEM sample holder in order to perform in situ measurements. This way, TEM annealed samples closely resembled the analogous transport samples. Moreover, this thin capping layer prevented any contamination of the a-NbSi film during the annealing, which ranged from 70$\,^{\circ}{\rm C}$ to 700$\,^{\circ}{\rm C}$. The annealing took place inside the vacuum chamber and lasted 15 minutes. Since the characteristic time for annealing-induced relaxation is smaller than 1 minute \cite{Querlioz2005, Myers2012, Lesueur1985these}, this duration is sufficient to obtain the same microscopic state of the material as for the transport samples.

Transport measurements were carried out down to 10 mK in a dilution refrigerator, using a TRMC2 resistance measurement bridge and standard AC lock-in detection techniques. The current applied to the sample was well within the linear regime of the $I$-$V$ characteristics for the considered films. All electrical leads were filtered from RF at room temperature.

%?????????????????????????????????????????????????????????????????????????????????????

%?????????????????????????????????????????????????????????????????????????????????????
\section{Structure of a-Nb$_{x}$Si$_{1-x}$ thin films}
\label{sec:Morpho}
%?????????????????????????????????????????????????????????????????????????????????????

The characterization of disordered thin film morphology is especially relevant for the study of the SIT, since the physical mechanisms explaining the destruction of superconductivity may differ for granular and morphologically homogeneous systems \cite{Feigelman2010, Valles1992}. We have therefore conducted AFM and TEM measurements on our as-deposited a-NbSi films in order to verify their continuity, amorphousness, and non-granularity (Sec. \ref{subsec:Morpho}) down to $d$ = 25 {\AA}.

Since annealing plays an important role in the present work, we have extended TEM measurements to films that were annealed in situ from 70$\,^{\circ}{\rm C}$ to 700$\,^{\circ}{\rm C}$ (Sec. \ref{subsec:Morpho_recuit}). We will show that we observe, through this technique, no structural change in a-NbSi films up to an annealing temperature of 500$\,^{\circ}{\rm C}$.

%//////////////////////////////////////////////////////////////////////////////////
\subsection{Morphology of as-deposited films}
\label{subsec:Morpho}

%............................................................................
\subsubsection{Transport measurements}

Before describing the microscopy measurements we have performed, we would like to emphasize that transport measurements are very sensitive to the sample structure and can thus give us some information on the microscopic nature of a-NbSi.

\begin{figure}[h]
\begin{centering}
\includegraphics[width=0.9\columnwidth]{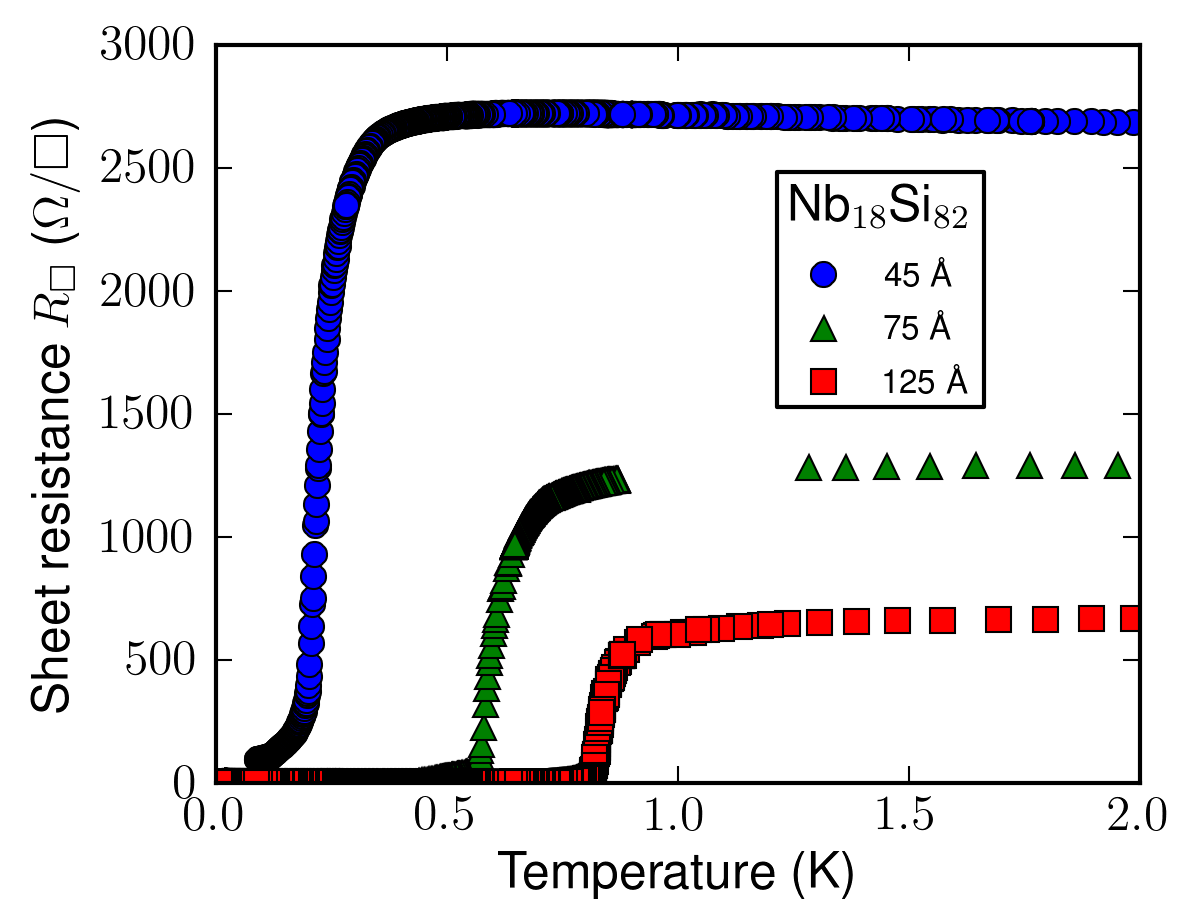}\\
  \caption{Sheet resistance as a function of temperature for Nb$_{18}$Si$_{82}$ samples of thicknesses ranging from 45 {\AA} to 125 {\AA}. The 25 {\AA}-thick sample of the same composition was found to be insulating.}
  \label{fig:transport}
\end{centering}
\end{figure}

Figure \ref{fig:transport} shows the low-temperature characteristics of three Nb$_{18}$Si$_{82}$ films for thicknesses of 125 {\AA}, 75 {\AA}, and 45 {\AA}. As has been previously measured in other a-NbSi samples \cite{Marrache2008}, the sheet resistance $R_\square(T)$ shows no sign of reentrance as usually observed for granular systems \cite{Haviland1989} when the thickness is lowered. Moreover, the superconducting transitions are a few tens of mK sharp, as is usual for homogeneous films.

We have also measured the electron-phonon coupling constant to be of $g_{e-ph}\simeq 50$ W.K$^{-5}$.cm$^{-3}$ at 100 mK, which is very similar to the values obtained in bulk metallic a-NbSi \cite{Marnieros2000}. This is another indication that all the film volume participates to the conduction and that there are no percolating structures.

%............................................................................

%............................................................................
\subsubsection{AFM measurements}

We have performed AFM measurements on films of thicknesses down to 25 {\AA}. All measured films were continuous, on a macroscopic scale, within the precision of the apparatus ($\simeq$ 20 nm in the sample plane). As we will see in the next paragraph, TEM observations will confirm this.

The sample surface was moreover found to be very smooth, with a RMS roughness limited by the AFM resolution ($\simeq$ 1 {\AA} in height). This is yet another strong indication of the non-granularity of a-NbSi films down to these thicknesses.
%............................................................................

%............................................................................
\subsubsection{TEM measurements}

\begin{figure}[h!]
\begin{centering}
\includegraphics[width=0.95\columnwidth]{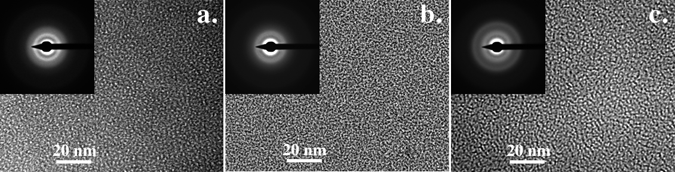}\\
  \caption{TEM images of a. the SiO$_2$ membrane that serves as substrate to TEM measurements ; b. 25 {\AA} and c. 100 {\AA}-thick Nb$_{18}$Si$_{82}$ film deposited onto the membrane. In the insets are the corresponding electronic diffraction patterns.}
  \label{fig:TEM}
\end{centering}
\end{figure}

TEM measurements were destined to confirm, with a better spatial resolution, the AFM results, namely the continuity and homogeneity of our films. For practical purpose, as stated in Sec. \ref{sec:synthesis}, the a-NbSi films were evaporated onto SiO$_2$ membranes for these observations.
%This difference of substrate with the transport samples should not notably modify the films morphology. Indeed, a-NbSi films deposited on thin membranes have previously been studied by transport measurements and have shown the same temperature dependence of the resistance as analogous films deposited onto bulk substrates \cite{Marnieros2013}.

Figure \ref{fig:TEM}.b (resp. \ref{fig:TEM}.c) shows a TEM image, along with the corresponding diffraction pattern, of a 25 {\AA} (resp. 100 {\AA})-thick Nb$_{18}$Si$_{82}$ films deposited onto such a SiO$_2$ membrane. Samples have been imaged at different positions, and no noticeable spatial difference either in the TEM images or the diffraction spectra was observed.

Due to the grain-like structures present on the TEM image of the commercial SiO$_2$ membrane (figure \ref{fig:TEM}.a), which are most probably due to mechanical stress effects, we estimate that we cannot visualize structures in a-NbSi films smaller than 20 {\AA} in size. However, the TEM resolution is good enough for us to be confident that our films contain no discontinuities or granular structures larger than this limit, as confirmed by transport measurements (sec. \ref{sec:Annealed}).

Besides, the diffraction spectrum for the 100 {\AA}-thick sample attests to the amorphousness of this film. In the case of the 25 {\AA}-thick film, the diffraction spectrum is more difficult to interpret due to the ratio of the film thickness to the membrane thickness, but it is reasonable to assume that films thinner than 100 {\AA} do not contain any crystallites, since they would also have been imaged in the 100 {\AA}-thick sample. This will also be confirmed by conductivity measurements (section \ref{subsec:conductivity}).

%............................................................................

%............................................................................
\subsubsection{Composition homogeneity}

\begin{figure}[h!]
\begin{centering}
\includegraphics[width=\columnwidth]{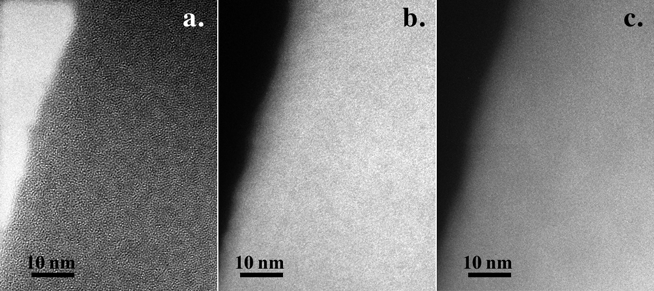}\\
  \caption{a. TEM image of a 25 {\AA}-thick Nb$_{18}$Si$_{82}$ film deposited onto a SiO$_2$ membrane. The image is taken near a membrane edge (left). b. EFTEM image of the same region of the sample for Nb 34 eV M X-ray and c. for Si 99 eV L X-ray.}
  \label{fig:EFTEM}
\end{centering}
\end{figure}

In order to check the uniformity of the film composition along the sample and to confirm its homogeneity in thickness, we have performed EFTEM measurements on the 25 {\AA}-thick Nb$_{18}$Si$_{82}$ TEM sample. We have considered the 34 eV and 99 eV energy edges for Nb and Si respectively, using a 20 eV slit. The results are shown figure \ref{fig:EFTEM}. Any local difference on the EFTEM signal can be attributed to a change in niobium concentration or in the film thickness. Each measurement was performed for different focuses and no visible structure remained, leading us to conclude the samples are very homogeneous both in composition and thickness. We estimate that any difference in the number of Nb atoms larger than $\pm$ 0.1\% can be detected by this experimental set-up with a spatial resolution of 10 {\AA}.

%............................................................................

%//////////////////////////////////////////////////////////////////////////////////

%//////////////////////////////////////////////////////////////////////////////////
\subsection{Annealed films}
\label{subsec:Morpho_recuit}

As previously stated, the present work focuses on the SIT in \emph{amorphous} NbSi films. Exposed to sufficiently high temperature, bulk ($d >$ 500 {\AA}) NbSi is known to crystallize \cite{Nava1986, Querlioz2005}. In order to use annealing as a relevant tuning parameter for our system, we therefore first had to establish the maximum temperature at which the amorphous character of these thin films is preserved.

For this, we have performed TEM and diffraction measurements on a 25 {\AA}-thick Nb$_{18}$Si$_{82}$ sample, that was annealed in situ at regularly-spaced temperatures ranging from 70$\,^{\circ}{\rm C}$ to 700$\,^{\circ}{\rm C}$.

\begin{figure}[h]
\begin{centering}
\includegraphics[width=\columnwidth]{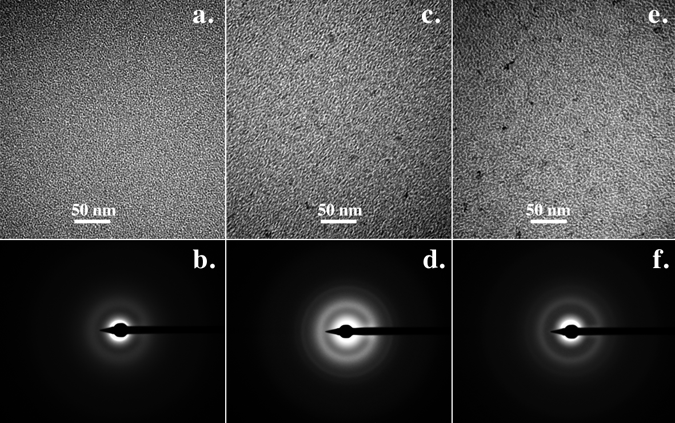}\\
  \caption{TEM images and diffraction patterns of a 25 {\AA}-thick Nb$_{18}$Si$_{82}$ sample after a 200$\,^{\circ}{\rm C}$ annealing (a. and b.), after a 500$\,^{\circ}{\rm C}$ annealing (c. and d.), and after a 700$\,^{\circ}{\rm C}$ annealing (e. and f.).}
  \label{fig:recuit}
\end{centering}
\end{figure}

The results are summarized figure \ref{fig:recuit}. We have found no significant change in the film structure until an annealing temperature of 500$\,^{\circ}{\rm C}$, consistant with the results obtained for bulk samples \cite{Querlioz2005}. Above this value, TEM images show the emergence of regions with a higher electron density. These have typical sizes of 50$\sim$100 {\AA}. Moreover, clearly defined fine spotless rings appear in the diffraction pattern. These signal the appearance of small cubic face centered structures (lattice parameter $a \simeq$ 4.31 {\AA}) compatible with the formation of partially crystallized nanometric particules of Nb$_3$Si ($a$ = 4.20 {\AA}). As the annealing temperature is further increased, the crystallites grow in size, but no additional feature appears on the diffraction pattern.

Considering our experimental resolution, we can be confident that no crystallites over 20 {\AA} in size appear before an annealing temperature of 500$\,^{\circ}{\rm C}$ or else they would have been imaged. However, because we deal with very thin samples and very few Nb atoms, we cannot rule out that precursors to this crystallization, of size smaller than 20 {\AA}, appear at lower annealing temperatures. This point will be further discussed in section \ref{subsec:conductivity}.
%//////////////////////////////////////////////////////////////////////////////////
%?????????????????????????????????????????????????????????????????????????????????????

%?????????????????????????????????????????????????????????????????????????????????????
\section{Effect of annealing on the transport properties of a-NbSi}
\label{sec:Annealed}
\begin{figure}[h!]
\begin{centering}
\includegraphics[width=0.75\columnwidth]{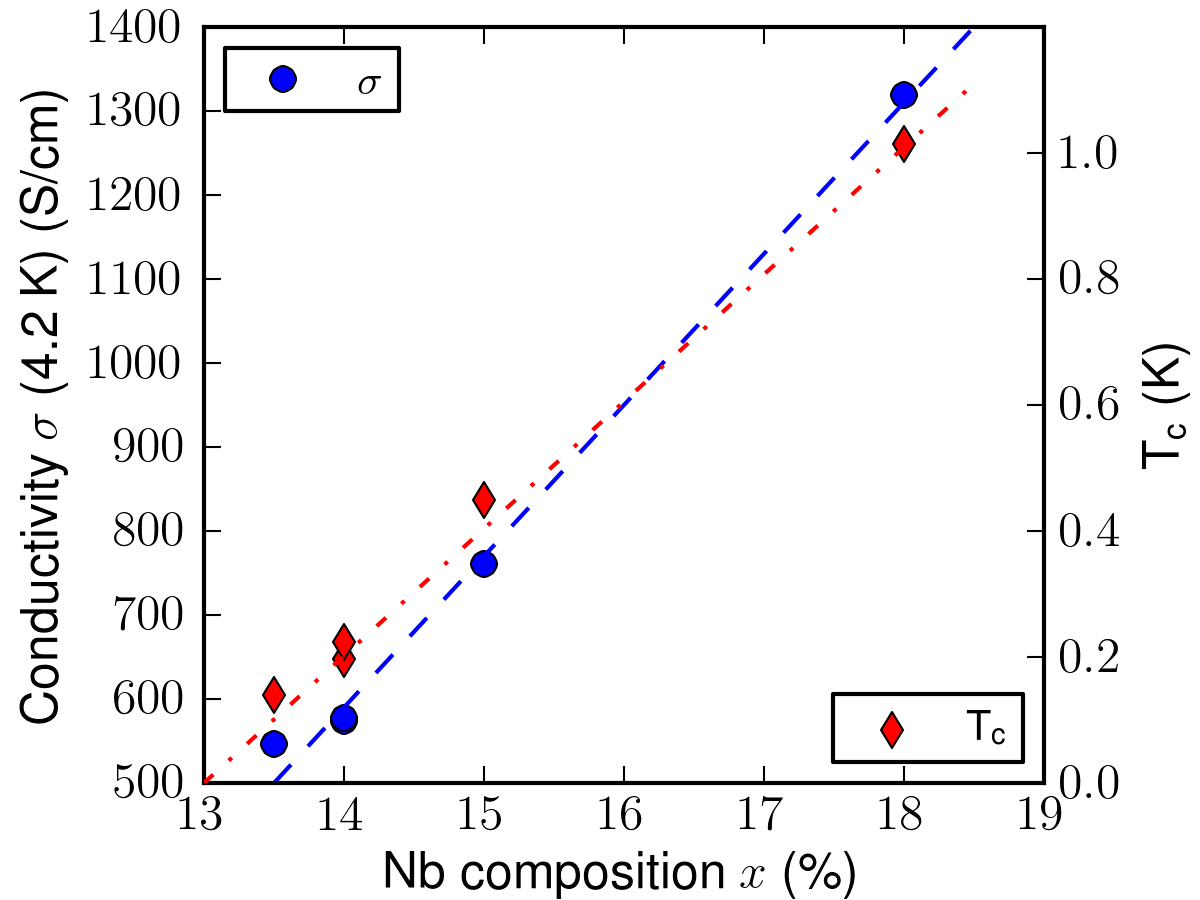}\\
  \caption{Evolution of the 4.2 K conductivity $\sigma_{4K}$ (blue dots) and of the $T_c$ (red diamonds) with the alloy composition $x$. The lines are guides to the eye.}
  \label{fig:pptes_bulk}
\end{centering}
\end{figure}

%?????????????????????????????????????????????????????????????????????????????????????
As-deposited a-NbSi films can exhibit superconducting, metallic or insulating behaviors, depending, notably, on the niobium content $x$ \cite{Dumoulin1993, Helgren2001, Lee1985}. In particular, for $x\gtrsim$12\%, a-NbSi films are superconducting. As an illustration, the evolutions of the 4K conductivity $\sigma$ and of the superconducting critical temperature $T_c$ with $x$ are given figure \ref{fig:pptes_bulk} for \emph{bulk} samples. As can be seen, for 3D a-NbSi films, the conductivity and the $T_c$ increase linearly with $x$, in accordance with previous studies on a-NbSi near the Metal-to-Insulator Transition \cite{Bishop1985}. This dependence is known to reflect the change in density of states induced by a change in Nb composition.

To study the effect of the annealing temperature on the superconducting properties of two-dimensional a-NbSi thin films, we have considered samples of two different niobium compositions, $x$=13.5\% and $x$=18\%, allowing us to span the portion of the phase diagram where $0.14\leq T_{c, bulk}\leq 1$ K. The superconducting films have thicknesses ranging from 45 {\AA} to 500 {\AA}. We can estimate the superconducting coherence length $\xi_{SC}\simeq\sqrt{\frac{0.18\hbar v_Fl}{k_BT_c}}$, to be of the order of 260 to 930 {\AA} depending on the film $T_c$, taking the mean free path to be the interatomic distance $l=2.65$ {\AA} \cite{Hucknall1992} and $v_F\simeq2.10^8$ cm.s$^{-1}$ \cite{Marnieros1998}, in excellent agreement with the estimate of the Ginzburg-Landau coherence length obtained by Nernst effect in films of similar composition \cite{Pourret2006, Pourret2007}. Except for the 500 {\AA}-thick films, the films all have $d<\xi_{SC}$, as can be seen from table \ref{tab:cond_unannealed}, which is the commonly accepted criterion for 2D superconductivity \cite{Feigelman2012}. The transport properties of these films have been studied at low temperature, for annealing temperatures, $\theta_a$ ranging from room temperature to 250$\,^{\circ}{\rm C}$.

\begin{figure}[h!]
\begin{centering}
\includegraphics[width=0.75\columnwidth]{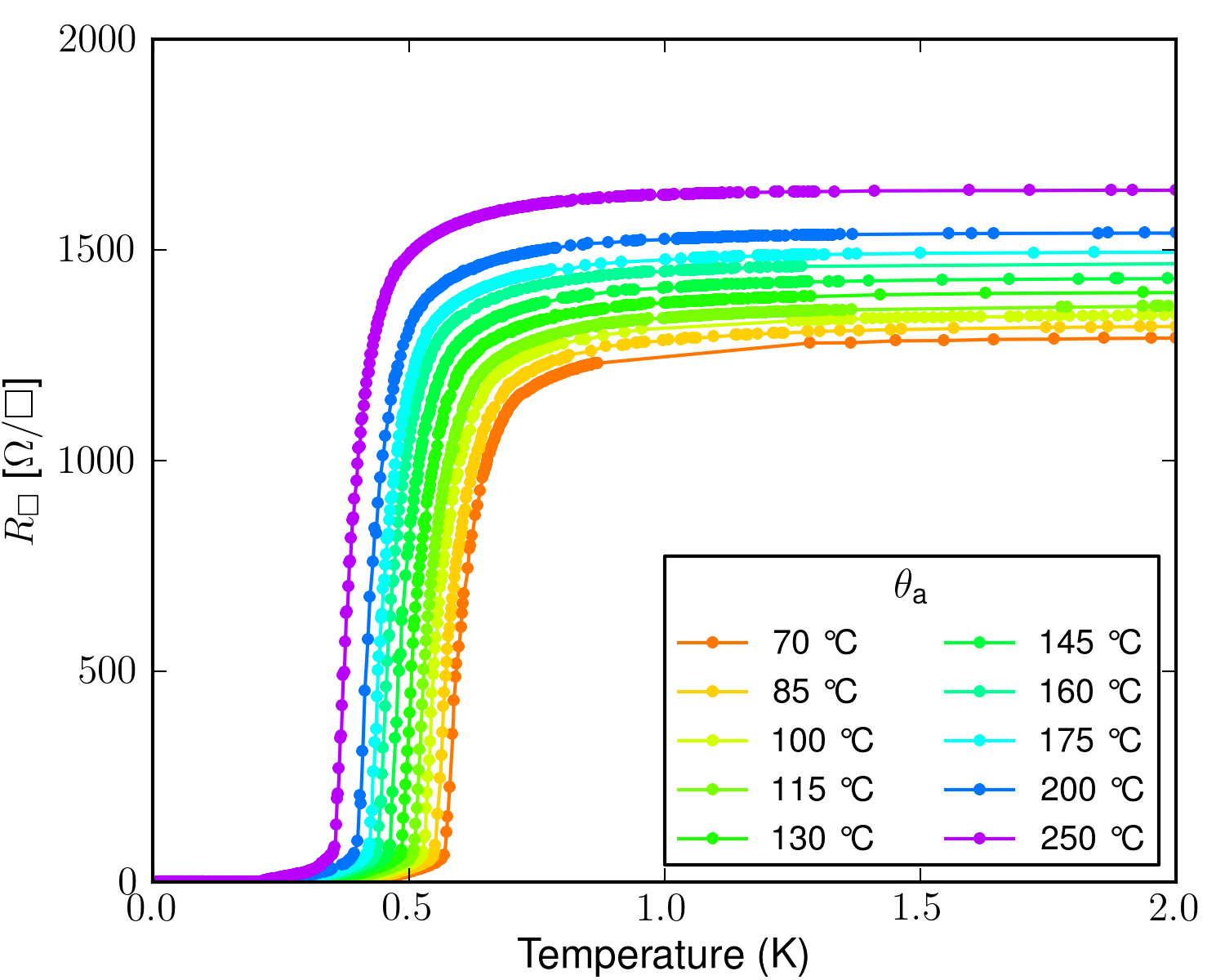}\\
  \caption{Evolution of the low temperature transport properties of a 75 {\AA}-thick Nb$_{18}$Si$_{82}$ film for $70\,^{\circ}{\rm C}<\theta_a<250\,^{\circ}{\rm C}$.}
  \label{fig:effet_annealing}
\end{centering}
\end{figure}

In this work, we have focused on the effect of moderate annealing on a-NbSi films. It is important to restate that, in the range of $\theta_a$ considered, the samples remain amorphous. We therefore do not consider the situation where the annealing induces partial or total crystallization of the film. Within this regime, the effect of increasing $\theta_a$ on superconducting a-NbSi films is twofold: the normal resistance increases and the $T_c$ decreases, as is shown for a typical sample figure \ref{fig:effet_annealing}. In the following sections we will detail these effects.

%//////////////////////////////////////////////////////////////////////////////////
\subsection{Evolution of the films conductivity}
\label{subsec:conductivity}

\begin{table}
\caption{4.2 K conductivity $\sigma$ and superconducting coherence length $\xi_{SC}$ of the different as-deposited samples. (N.S. = Non-Superconducting).}
\label{tab:cond_unannealed}
\begin{tabular}{|c|c|c|c|c|}
\hline
$x$ (\%) & $d$ ({\AA}) & $\sigma$ ($\times 10^{2}$ S.cm$^{-1}$) & $T_c$ (K) & $\xi_{SC}$ ({\AA})\\
\hline
\hline
13.5 & 50 & 3.1 & N.S.  & \\
13.5 & 75 & 4.2 & N.S.  & \\
13.5 & 100& 4.6 & N.S.  & \\
13.5 & 200& 5.4 & N.S.  & \\
13.5 & 250& 5.4 & 0.080 & 930\\
13.5 & 300& 5.6 & 0.104 & 815\\
13.5 & 500& 5.5 & 0.140 & 703\\
\hline
18 & 20 & 0.91 & N.S.   &\\
18 & 25 & 3.8 & N.S.    &\\
18 & 30 & 5.5 & N.S.    &\\
18 & 45 & 8.4 & 0.196   & 594\\
18 & 75 & 10.3 & 0.568  & 349\\
18 & 125& 11.9 & 0.724  & 309\\
18 & 125& 12.2 & 0.809  & 292\\
18 & 500& 13.2 & 1.015  & 261\\
\hline
\end{tabular}
\end{table}

For thick samples ($d >$ 150 {\AA}), the conductivity of as-deposited a-NbSi only depends on the alloy composition. For thinner films, the conductivity decreases with the thickness (table \ref{tab:cond_unannealed}), as could be expected from Fuchs law \cite{Chopra1969} combined with localization and interaction effects \cite{Graybeal1984}.

\begin{figure}[h!]
\begin{centering}
\includegraphics[width=0.85\columnwidth]{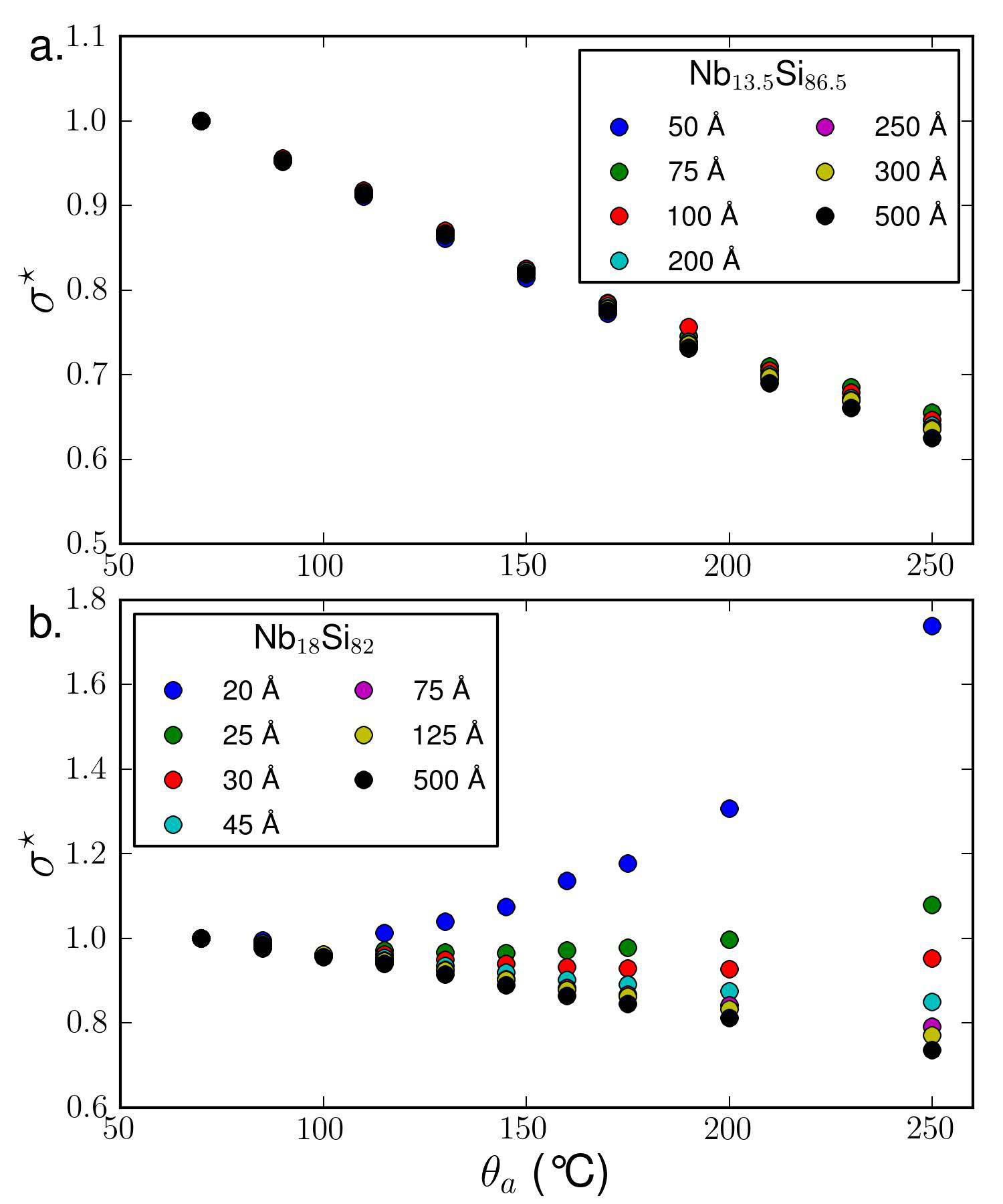}\\
  \caption{Evolution of the normalized 4.2 K conductivity $\sigma^*_{4K} (\theta_a)= \frac{\sigma_{4K}(\theta_a)}{\sigma_{4K}(\theta_a=70\,^{\circ}{\rm C})}$ with the annealing temperature for the Nb$_{13.5}$Si$_{86.5}$ samples (a.) and the Nb$_{18}$Si$_{82}$ samples (b.).}
  \label{fig:conductivite_renorm}
\end{centering}
\end{figure}

When moderately annealed, the conductivity of a-NbSi decreases almost linearly with the annealing temperature $\theta_a$ and $\sigma^*_{4K} (\theta_a)= \frac{\sigma_{4K}(\theta_a)}{\sigma_{4K}(\theta_a=70\,^{\circ}{\rm C})}$ is a function independent of the film thickness at fixed composition (figure \ref{fig:conductivite_renorm}). This linear behavior of $\sigma^*_{4K}$ with $\theta_a$ gives us ground to consider $\theta_a$ as a characterization of the disorder level in the sample, albeit not being an intrinsic relevant physical parameter \cite{Lesueur1985}. Moreover, from this linear relation between $\sigma^*_{4K}$ and $\theta_a$, we can verify that the effective annealing temperature of as-deposited films indeed is $70\,^{\circ}{\rm C}$.

Strongly annealed samples deviate from this trend: when a film is annealed over a critical temperature $\theta_{a,c}(d)$, its conductivity increases with $\theta_a$. We interpret this to be the onset of compositional ordering which eventually leads to crystallization, as has been observed in ref. \cite{Nava1986}. We have seen no sign of crystallites by TEM measurements (\ref{subsec:Morpho_recuit}) which leads us to think that these crystallites, if any, are small ($<$ 20 {\AA}). Thinner films are, for geometrical reasons, more sensitive to the formation of a small portion of a more metallic phase, cristalline or not, which could explain the thickness dependence of $\theta_{a,c}$.

These results show that transport measurements are extremely sensitive to the sample microscopic structure. Any deviation from the common $\sigma^*_{4K}(\theta_a)$ law was therefore considered to be the indication that, at a given thickness, the sample structure moved away from amorphousness and/or composition homogeneity. In the rest of this paper, the corresponding data will thus not be taken into account.

%//////////////////////////////////////////////////////////////////////////////////

%//////////////////////////////////////////////////////////////////////////////////
\subsection{Evolution of the superconducting properties}
\label{subsec:transport}

\begin{figure}[h!]
\begin{centering}
\includegraphics[width=0.85\columnwidth]{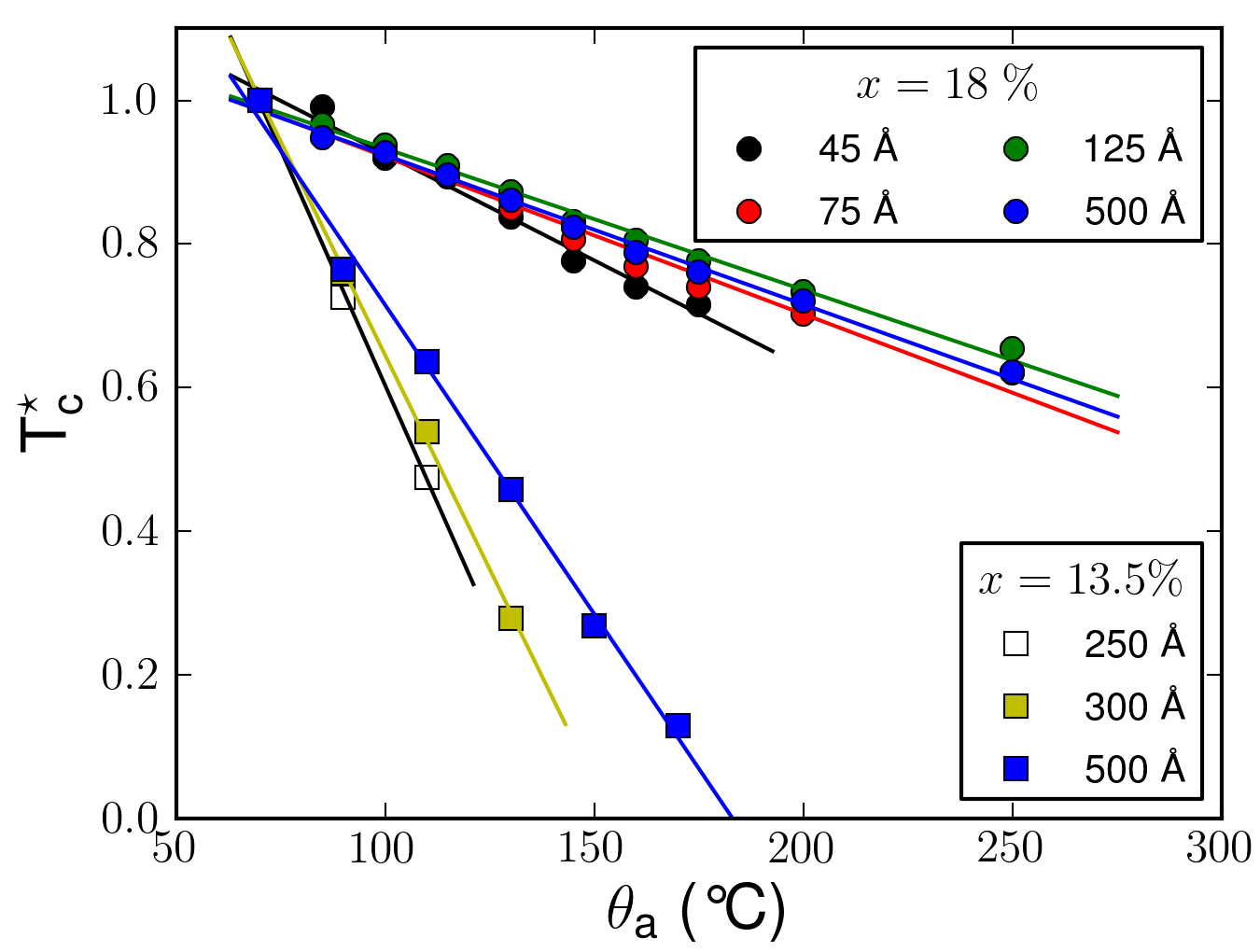}\\
  \caption{Evolution of the normalized superconducting critical temperature $T_c^*=\frac{T_c(\theta_a)}{T_c(\theta_a=70\,^{\circ}{\rm C})}$ with the annealing temperature for Nb$_{13.5}$Si$_{86.5}$ (squares) and Nb$_{18}$Si$_{82}$ (circles) samples.}
  \label{fig:Tc}
\end{centering}
\end{figure}

Figure \ref{fig:Tc} shows the evolution of $T_c^*=\frac{T_c(\theta_a)}{T_c(\theta_a=70\,^{\circ}{\rm C})}$ as a function of $\theta_a$. As mentionned earlier, increasing $\theta_a$ tunes the sample closer to the SIT and $T_c$ therefore is a decreasing function of $\theta_a$. Moreover, $|\frac{\text{d}T_c^*}{\text{d}\theta_a}|$ is smaller for large values of $x$: for samples with higher Nb content, the superconductivity is stronger and the effect of annealing is less felt.

%//////////////////////////////////////////////////////////////////////////////////
%?????????????????????????????????????????????????????????????????????????????????????

%?????????????????????????????????????????????????????????????????????????????????????
\section{Discussion on the effect of annealing}
\label{sec:Discussion}
%?????????????????????????????????????????????????????????????????????????????????????

%//////////////////////////////////////////////////////////////////////////////////
%\subsection{General framework of our interpretation}

Annealing of amorphous metallic alloys has been studied in the context of their interest as engineering materials \cite{Gilman1975, Kim2012, Legeay2012, Moon2002, Guillotel2007, Querlioz2005, Lesueur1985}. It has been shown that their properties, whether mechanical, magnetic or electrical could be altered even by relatively low temperature annealing which still preserved their amorphousness.

Although the study of microscopic changes induced by annealing is beyond the scope of this article, it should nonetheless be mentioned that the effects of such thermal treatments have been found to be threefold \cite{Egami1978b}. The first effect consists in diffusion processes. The other two effects can be attributed to structural relaxations, either chemical or topological. These are collective phenomena which redistribute quenched-in defects into lower energy configurations. Nominally, the material undergoes an extensive rearrangement of neighboring atoms without change in the mean nearest neighbor distance \cite{Egami1978a}. The structural relaxations can either be compositional - introducing chemical short range ordering (CSRO) - or topological - introducing topological short range ordering (TSRO) where the relative positions of the atoms are rearranged regardless of their chemical nature.

In the case of a-NbSi thin films, for the considered range of annealing temperatures ($\theta_a<250\,^{\circ}{\rm C}$), it is reasonable to rule out significant diffusion processes. Indeed, these have measurable effects only for annealing temperatures typically above 1000$\,^{\circ}{\rm C}$: in a compound similar to a-NbSi, the diffusion constant of an ion at 250 $\,^{\circ}{\rm C}$ has been measured to be of about 4.10$^{-19}$ cm$^2$.s$^{-1}$ \cite{Gupta1975}. For an annealing time of one hour, the effective length covered by the atoms can therefore be estimated to be of the interatomic distance, at most. Moreover, the diffusion of atoms from the SiO$_2$ under- and/or over-layers into the a-NbSi film of interest is negligible: the kinematics of oxygen diffusion in silicon are very slow \cite{Devine1993}. In addition, at temperatures above 1000$\,^{\circ}{\rm C}$, far above our maximum annealing temperature, silicon atoms tend to diffuse from Si layers into SiO$_2$ \cite{Tromp1985} rather than the contrary. Niobium atoms should also not be impacted by this very limited diffusion, thanks to their sparsity.

Secondly, as explained in the previous sections and verified in the literature \cite{Querlioz2005, Nava1986}, a-NbSi thin films remain amorphous, for $\theta_a<\theta_{a,c}$, and there is no sign of chemical segregation within the films. Given the compositions we are working with, it is therefore reasonable to assume that CSRO does not affect the films macroscopic properties, and only plays a non-negligible role for temperatures higher than $\theta_{a,c}$.

TSRO, however, might provide a plausible explanation for the effect of annealing in a-NbSi films. Indeed, these relaxation processes induce atomic movements on lengthscales a priori smaller than the inter-atomic distance. Beal and Friedel \cite{Beal1964} have theoretically studied the effect of such relaxation processes on the diffusion induced by pairs of scatterers and shown that this reorganization towards a local order always leads to an increase of the local resistivity. In systems close to the Metal-to-Insulator Transition, as is a-NbSi in these conditions, this effect can be dramatic since a small increase in local resistivity can translate into important effects on the macroscopic properties of the films \cite{bande_des_4_1979}.

Lastly, it is important to note that previous studies have concluded that annealing a-NbSi up to $\theta_a=260\,^{\circ}{\rm C}$ induces no change in the carrier density \cite{Nava1986}.

For the purpose of our analysis, we will limit ourselves to the macroscopic effects of these annealing treatments. As previously seen, annealing modifies the samples macroscopic conductivity. It can therefore be seen as a tuning parameter for disorder.

In disordered superconducting films, there has been numerous attempts to link the superconducting properties to the film disorder, measured by the product $k_Fl$. In two-dimensional systems, $R_\square=\frac{h}{e^2}\frac{1}{k_Fl}$, so that the square resistance has often been put forward as a way to quantify the mean sample disorder.

Theoretically, the Kosterlitz-Thouless critical temperature was predicted to be driven by $R_\square$ in disordered superconducting thin films \cite{Beasley1979, Halperin1979, Fiory1983}. In Maekawa and Fukuyama's theory explaining the initial effects of disorder on the $T_c$ \cite{Maekawa1982} through a purely fermionic scenario implying an increase of Coulomb interactions as $R_\square$ grows larger, $\frac{\Delta  T_c}{T_c}$ is a function of $R_\square$ as well. The extension of this theory by Finkelstein \cite{Finkelstein1994} also relates a reduction in $T_c$ to the sole $R_\square$ for a given compound. Another scenario, of a bosonic nature, predicts a change in ground state from superconducting to insulating at large disorder for an universal critical sheet resistance $R_{\square,c}$ \cite{Fisher1990bis}, which is temperature independent within the quantum critical region. The close link between a film sheet resistance and its critical temperature has also been underlined in recent theoretical reviews \cite{Larkin1999, Gantmakher2010}.

Numerous experiments have confirmed the correlation between $T_c$ and $R_\square$ in very different disordered systems: InOx \cite{Fiory1984}, a-Ga \cite{Jaeger1986}, MoC \cite{Lee1990}, and NbN \cite{Ezaki2012} for example. In these experiments, the disorder was made to vary through a change in the film thickness. Even more convincing were the results grouping different compounds or different ways of tuning the disorder, which showed a strikingly similar destruction of superconductivity when $R_\square$ grew: Strongin et al. \cite{Strongin1970} showed the same $T_c$ reduction with $R_\square$ for Bi and Pb films ; Valles et al. \cite{Valles1989} did so for Pb and Sn films ; Graybeal et al. \cite{Graybeal1984, Graybeal1985} presented similar results for a-MoGe films of different compositions and thicknesses. Shahar et al. \cite{Shahar1992} also showed evidence for a well-defined $T_c(R_\square)$ relationship for InOx films of different oxygen contents.

In the following, we will show that our results are in contradiction with these previous studies, and that the reduction of the film thickness does not affect the superconducting properties in the same way as the composition $x$ or the annealing temperature $\theta_a$ does. In other words, $R_\square$ is not, in our system, a relevant parameter to describe the superconductor-to-insulator transition.

%//////////////////////////////////////////////////////////////////////////////////

%//////////////////////////////////////////////////////////////////////////////////
\subsection{Comparison between the effects of $x$ and $\theta_a$}
\label{subsec:comp_x_rec}

\begin{figure}[h!]
\begin{centering}
\includegraphics[width=0.95\columnwidth]{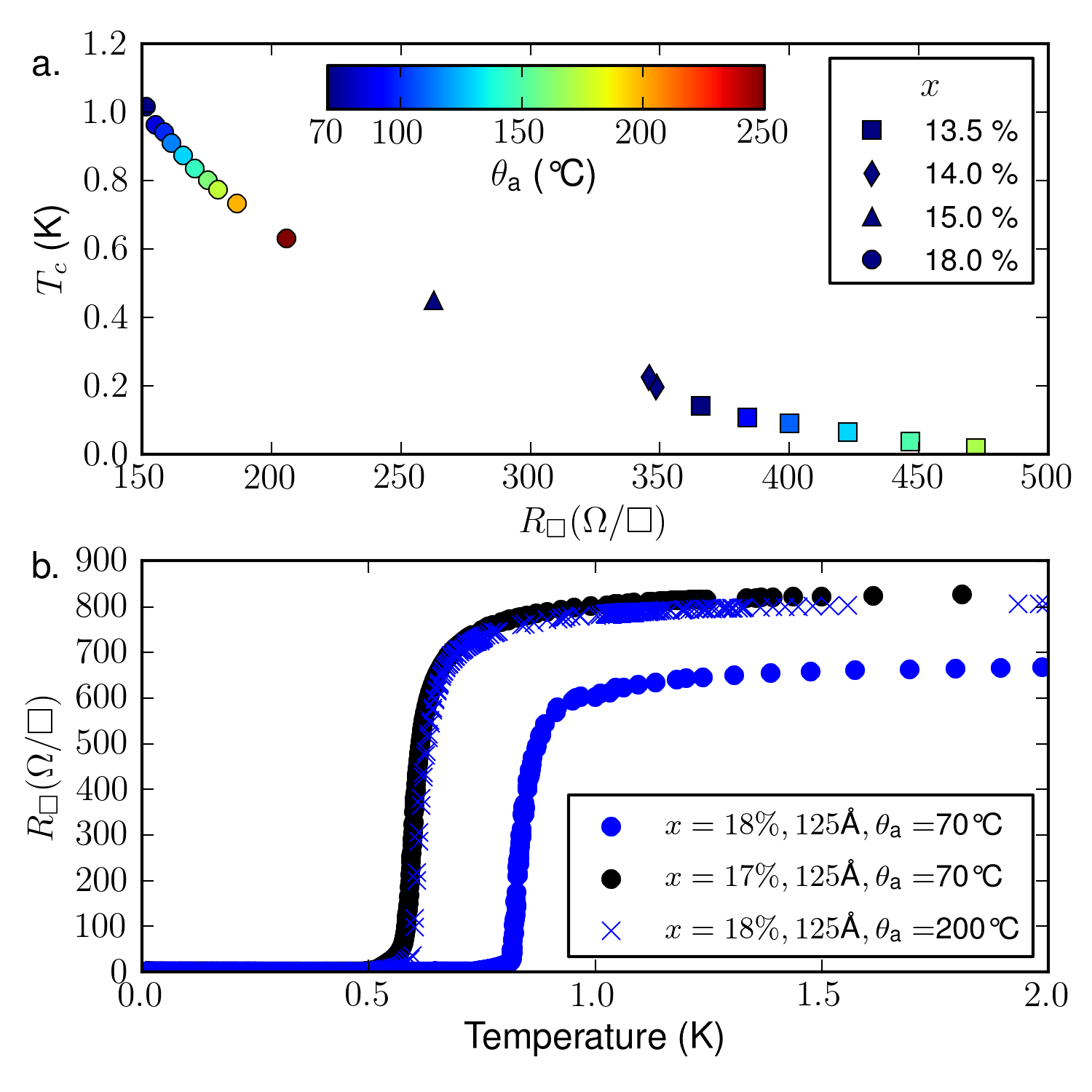}\\
  \caption{a. $T_c$ as a function of the disorder parameter $R_\square$ for several 500 $\AA$ thick samples. Each symbol represents one composition and each color one annealing temperature. The samples of $x=14\%$ and $x=15\%$ have been previously studied \cite{Marrache2008, Crauste2011} and are given for comparison. b. Annealing and a change in composition affect the superconducting properties in the same way: starting from a 125 $\AA$ thick Nb$_{18}$Si$_{82}$ sample (blue circles), the superconductivity can be weakened in the same manner ($T_c$ and $R_\square$) by annealing (blue crosses) or by a change of composition (red circles).}
  \label{fig:x_recuit}
\end{centering}
\end{figure}

Annealing and the film composition both modify the disorder parameter $R_\square\propto\frac{1}{k_Fl}$: the film composition influences the density of states and hence modifies $k_F$, whereas annealing changes the effective mean free path $l$ through a variation of local conductivity. It can therefore be expected that, at fixed thickness $d$, these two tuning parameters have the same influence on the destruction of superconductivity. This effect can be seen on figure \ref{fig:x_recuit}.a: the depletion of $T_c$ relates one-to-one with the square resistance $R_\square$ as the SIT is approached either through annealing (each color symbolizing one annealing temperature) or through a composition change (each composition being represented by a different symbol).

The fact that, at a given sample thickness, $T_c$ is entirely determined by $R_\square$ can also be seen in figure \ref{fig:x_recuit}.b: starting from a 125 $\AA$ thick Nb$_{18}$Si$_{82}$ sample, one can achieve the same reduction in $T_c$, \emph{and the same $R_\square$} with a $\theta_a=200\,^{\circ}{\rm C}$ annealing \emph{or} with a reduction of the composition to $x=17\%$.

At fixed thickness, $R_\square$ can therefore be considered to be a relevant parameter to describe the SIT.
%//////////////////////////////////////////////////////////////////////////////////

%//////////////////////////////////////////////////////////////////////////////////
\subsection{Comparison between the effects of $d$ and $\theta_a$}
\label{subsec:comp_d_rec}

\begin{figure}[h!]
\begin{centering}
\includegraphics[width=0.95\columnwidth]{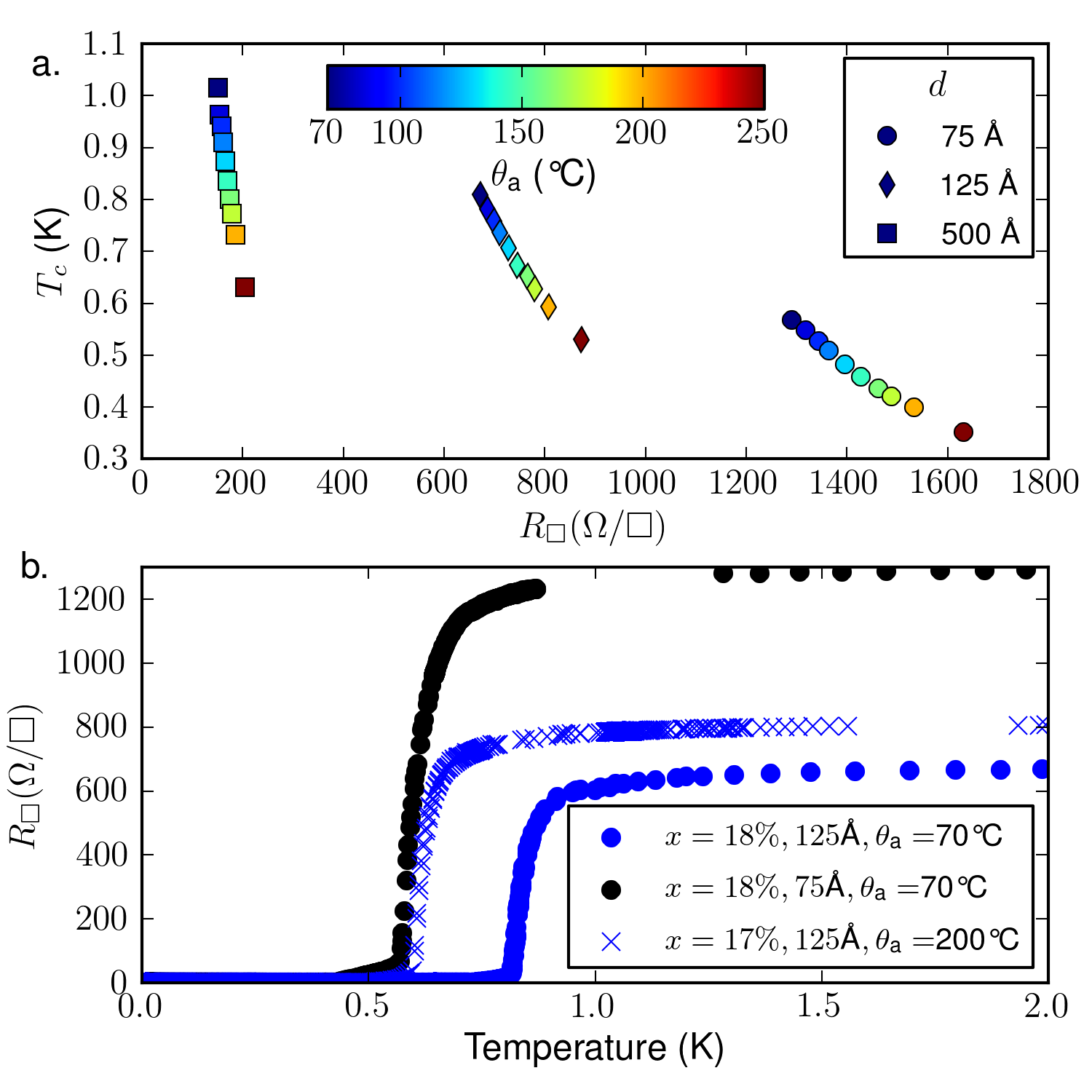}\\
  \caption{a. $T_c$ as a function of the disorder parameter $R_\square$ for several Nb$_{18}$Si$_{82}$ samples. Each symbol represents one thickness and each color one annealing temperature. b. Annealing and a change in thickness affect the superconducting properties in different ways: starting from a 125 $\AA$ thick Nb$_{18}$Si$_{82}$ sample (blue circles), $T_c$ can be weakened by annealing (blue crosses) or by a reduction of thickness (red circles). The corresponding square resistances are however different.}
  \label{fig:d_recuit}
\end{centering}
\end{figure}

The comparison between the effects of annealing with those induced by a reduction of the sample thickness is however more difficult: how a change in $d$ affects $R_\square$ is unclear. In a similar manner as previously, let us compare the effects of both parameters on $T_c$. As can be seen figure \ref{fig:d_recuit}.a, there is \emph{no} one-to-one relationship between $T_c$ and $R_\square$: at a given composition $x$, the $T_c$ reduction induced by the annealing $|\frac{\text{d}T_c}{\text{d}\theta_a}|$ is much more important than the thickness-induced effect ($|\frac{\text{d}T_c}{\text{d}d}|$).

Again, figure \ref{fig:d_recuit}.a shows that, starting from a 125 $\AA$ thick Nb$_{18}$Si$_{82}$ sample, one can achieve the same reduction in $T_c$ with a $\theta_a=200\,^{\circ}{\rm C}$ annealing or a reduction of the thickness to $d=75\AA$. However, the corresponding $R_\square$ differ by a factor of almost 2 in the two cases. In other words, the knowledge of $R_\square$ is not sufficient to predict the superconducting behavior of the films when $d$ is made to vary.
%/////////////////////////////////////////////////////////////////////////////////

%//////////////////////////////////////////////////////////////////////////////////
\subsection{Quantifying disorder}
\label{subsec:kFl}

\begin{figure}[h!]
\begin{centering}
\includegraphics[width=0.75\columnwidth]{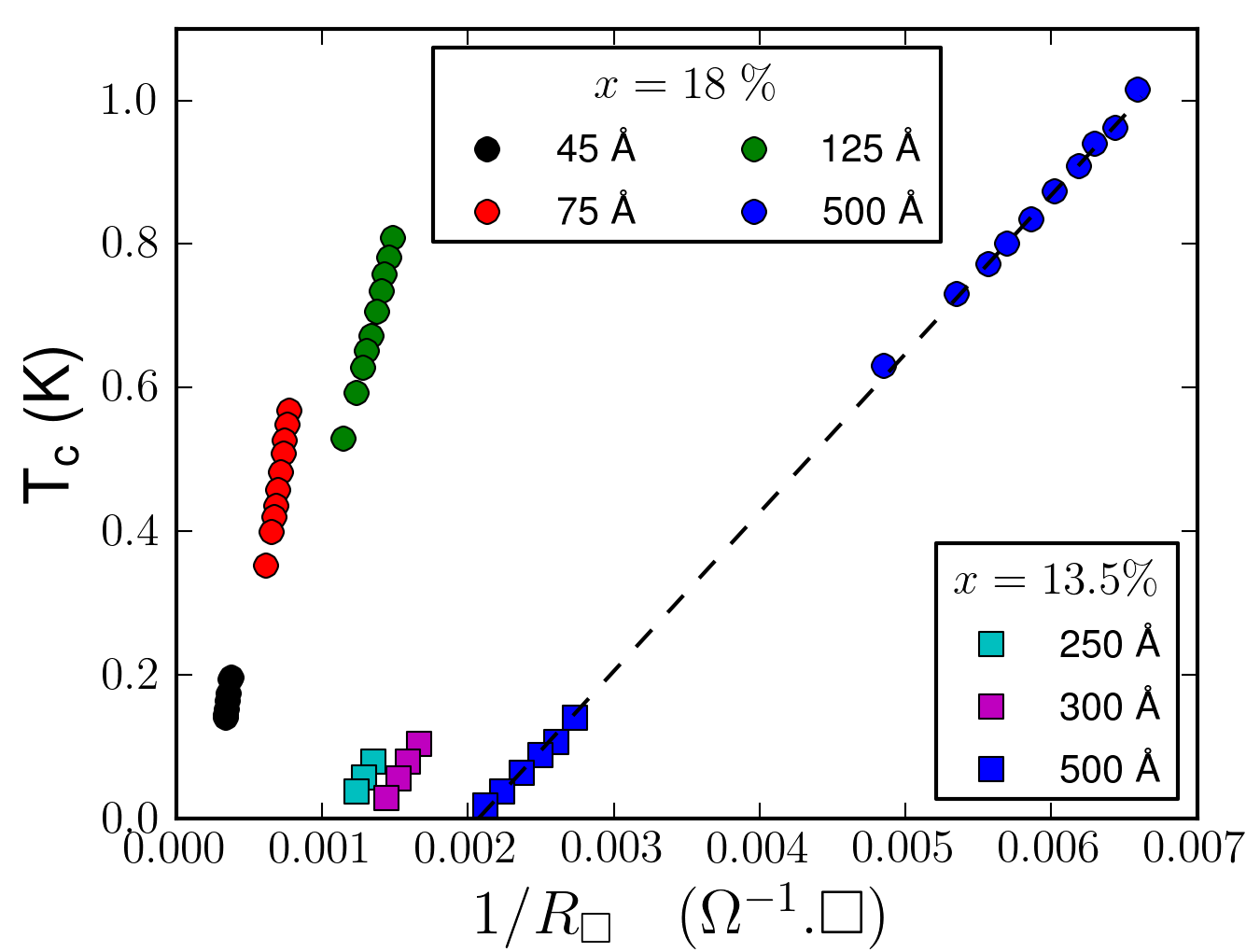}\\
  \caption{$T_c$ as a function of $R_\square$ for all films studied. Films of a given thickness follow the same trend regarding the suppression of superconductivity (black dashed line as guide to the eye).}
  \label{fig:T_c_kFl}
\end{centering}
\end{figure}

The results of the two previous sections are summarized figure \ref{fig:T_c_kFl}. The critical temperatures for all superconducting films are plotted with respect to the corresponding $\frac{1}{R_\square}$ for both compositions ($x=13.5\%$ and $x=18\%$) and for different annealing temperatures. Whereas samples of 500 {\AA} follow the same $T_c$ suppression law, as was suggested by \cite{Shahar1992}, the trend seems to be specific to each of the considered film thickness, and no universal behavior could be determined.

\begin{figure}[h!]
\begin{centering}
\includegraphics[width=0.75\columnwidth]{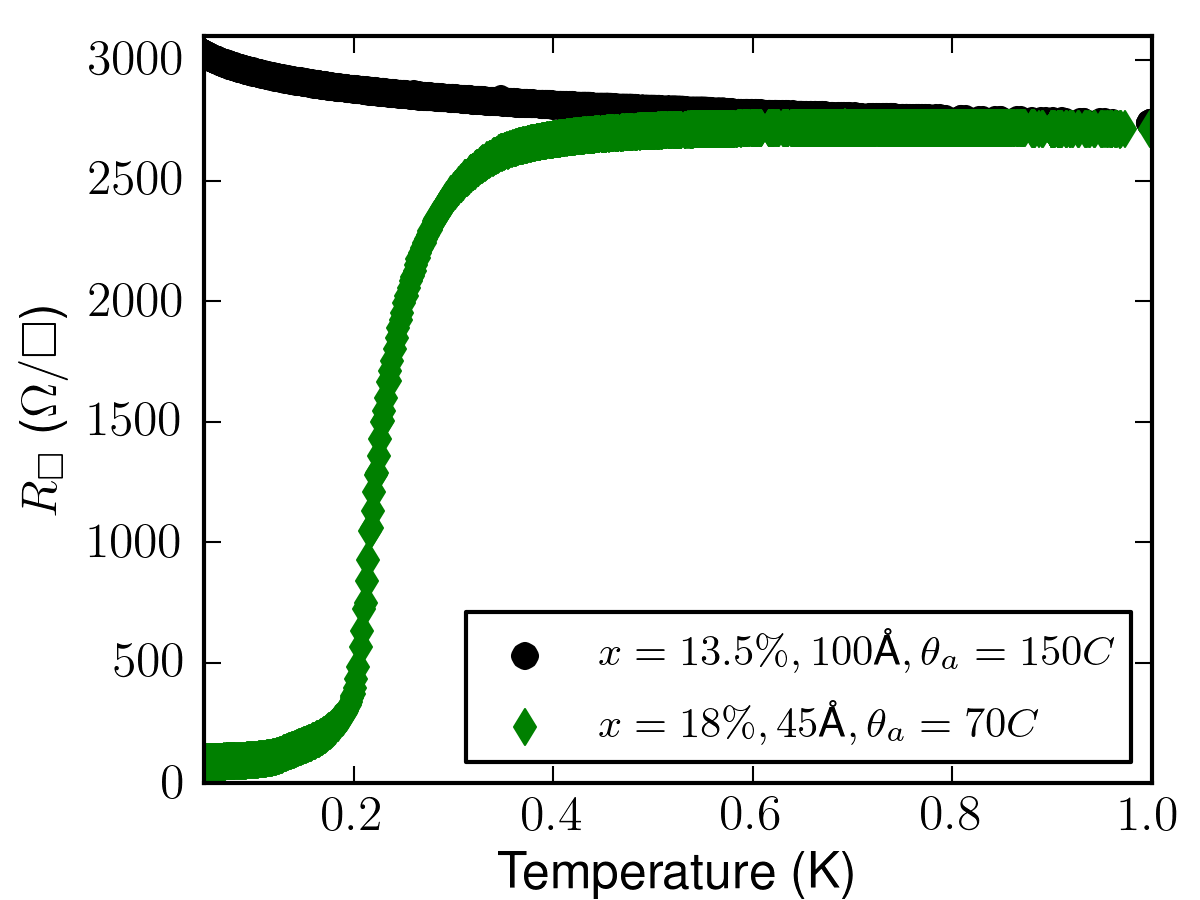}\\
  \caption{Transport properties of two films with the same $R_\square$. Despite having the same disorder parameter $\frac{1}{R_\square}$, these two films are on opposite sides of the SIT.}
  \label{fig:kFl}
\end{centering}
\end{figure}

As a prominent illustration of this point, two films of the same $R_\square$ have been found to be on opposite sides of the superconductor-to-insulator transition, as shown figure \ref{fig:kFl}. This contradicts the theoretical predictions that $R_\square$ solely determines the ground state near the SIT \cite{Fisher1990,Feigelman2010}.

%//////////////////////////////////////////////////////////////////////////////////

%?????????????????????????????????????????????????????????????????????????????????????

%?????????????????????????????????????????????????????????????????????????????????????
\section{Conclusion}
\label{sec:Conclusion}
%?????????????????????????????????????????????????????????????????????????????????????
We have studied the disorder-tuned suppression of superconductivity in a-NbSi films, through the variation of thickness, composition and annealing. We have shown that moderate annealing weakened superconductivity in these films, while the chemical composition of the films was unchanged and they remained amorphous. The suppression of superconductivity through a variation of composition or of annealing temperature could be explained, at a given thickness, through a single parameter $R_\square$. However, the effect of thickness reduction could not be satisfactorily explained this way. These arguments imply that $R_\square$ is not the relevant parameter to fully describe the disorder in the vicinity of a SIT. The transition tuned by the reduction of the sample thickness seems to play a special role, distinct from the effect of a modification of the microscopic disorder, here tuned by the annealing temperature. To the best of our knowledge, this difference remains to be understood.

Let us end with the mention that, for the different applications of superconducting films, it is particularly interesting to have an extrinsic parameter, like annealing, which could be used to fine-tune their properties, mainly the normal resistance $R_{\square,n}$ and the superconducting critical temperature $T_c$, after the thin film synthesis. We can cite the example of Transition Edge Sensors, which operating temperature could thus be finely adjusted to the need of the considered experiment \cite{Crauste2011}.
%**************************************************************************************************

% Acknowledgments *********************************************************************************

\begin{acknowledgments}
We gratefully thank Odile Kaitasov and Erwan Oliviero for their assistance in the TEM and EFTEM measurements. We acknowledge the help of Cyril Bachelet for RBS measurements. We are also indebted to Marie-Odile Ruault for many fruitful discussions. This work has been partially supported by the ANR (grants No. ANR-06-BLAN-0326 and ANR-2010-BLANC-0403-01), and by the Triangle de la Physique (grant No. 2009-019T-TSI2D).
\end{acknowledgments}
%**************************************************************************************************

%\newpage %Just because of unusual number of tables stacked at end

% Bibliographie ***********************************************************************************

\bibliography{Recuit_PRB}% Produces the bibliography via BibTeX.

%**************************************************************************************************

\end{document}